\documentclass[12pt]{article}

\usepackage[dvips]{graphicx}
\usepackage{lathuile}

\def\PRL                     {\mbox{Phys. Rev. Lett.}}
\def\PRD                     {\mbox{Phys. Rev. D}}

\newcommand{\Journal}[4]     {{#1} {\bf {#2}}, {#3} ({#4})}
\newcommand{\WWWAddr}[1]     {{\tt {#1}}}
\newcommand{\hepref}[2]      {{\tt hep-{#1}/{#2}}}

\newcommand{\FermiConf}[2]   {Fermilab-Conf-{#1}/{#2}-E}

\newcommand{\CERNEP}[2]      {CERN-EP-{#1}-{#2}}

\newcommand{\mygraphics}[1]{%
  \includegraphics[width=\linewidth,clip=false]{#1}\vspace{0pt}
}

\newcommand {\unitexp}[2] {\mbox{{#1}$^{\mathrm{#2}}$}}
\newcommand {\scinot}[2]  {\mbox{{#1}$\times$10$^{\mathrm{#2}}$}}

\newcommand {\order}[1]   {\mbox{$\mathcal{O}$(#1)}}

\newcommand {\ra}       {\mbox{$\rightarrow$}}
\newcommand {\Dzero}    {\mbox{D{\O}}}

\newcommand {\Zboson}   {\mbox{Z}}
\newcommand {\Wboson}   {\mbox{W}}
\newcommand {\Zz}       {\mbox{Z$^0$}}

\newcommand {\Hgeneral} {\mbox{$\phi$}}
\newcommand {\Hsm}      {\mbox{$h_{SM}$}}
\newcommand {\hlomssm}  {\mbox{$h$}}
\newcommand {\Hhimssm}  {\mbox{$H$}}
\newcommand {\Amssm}    {\mbox{$A$}}
\newcommand {\Hpmmssm}  {\mbox{$H^{\pm}$}}

\newcommand {\qrk}[1]   {\mbox{$#1$}}
\newcommand {\aqrk}[1]  {\mbox{$\bar{#1}$}}

\newcommand {\nutau}    {\mbox{$\nu_{\tau}$}}

\newcommand {\pbp}      {\mbox{$\bar{\mbox{p}}$p}}
\newcommand {\ppb}      {\mbox{p$\bar{\mbox{p}}$}}
\newcommand {\bbb}      {\mbox{\qrk{b}\aqrk{b}}}
\newcommand {\ccb}      {\mbox{\qrk{c}\aqrk{c}}}
\newcommand {\ttb}      {\mbox{\qrk{t}\aqrk{t}}}
\newcommand {\epem}     {\mbox{e$^+$e$^-$}}
\newcommand {\lplm}     {\mbox{$\ell^+ \ell^-$}}
\newcommand {\WW}       {\mbox{\Wboson \Wboson}}
\newcommand {\ZZ}       {\mbox{\Zboson \Zboson}}
\newcommand {\WZ}       {\mbox{\Wboson \Zboson}}

\newcommand {\ggH}      {\mbox{gg\ra \Hsm}}
\newcommand {\WH}       {\mbox{\Wboson \Hsm}}
\newcommand {\ZH}       {\mbox{\Zboson \Hsm}}

\newcommand {\tHb}      {\mbox{\qrk{t}\ra \Hpmmssm \qrk{b}}}
\newcommand {\phibb}    {\mbox{\Hgeneral \bbb}}

\newcommand {\Hbb}      {\mbox{\Hsm \ra \bbb}}

\newcommand {\lnbb}     {\mbox{$\ell \nu$\bbb}}
\newcommand {\llbb}     {\mbox{\lplm \bbb}}
\newcommand {\nnbb}     {\mbox{$\nu \bar{\nu}$\bbb}}
\newcommand {\jjbb}     {\mbox{$jj$\bbb}}
\newcommand {\llnn}     {\mbox{\lplm $\nu \bar{\nu}$}}
\newcommand {\lpmlpmjj} {\mbox{$\ell^{\pm} \ell^{\pm} jj$}}
\newcommand {\lpmlpmlmp}{\mbox{$\ell^{\pm} \ell^{'\pm} \ell^{\mp}$}}

\newcommand {\MHsm}    {\mbox{$M_{h}$}}
\newcommand {\Mhlo}    {\mbox{$M_{h}$}}
\newcommand {\MHhi}    {\mbox{$M_{H}$}}
\newcommand {\MA}      {\mbox{$M_{A}$}}
\newcommand {\MHpm}    {\mbox{$M_{H^{\pm}}$}}

\newcommand {\SUtwol}  {\mbox{$SU(2)_L$}}
\newcommand {\Uone}    {\mbox{$U(1)$}}

\newcommand {\Et}      {\mbox{$E_T$}}
\newcommand {\Pt}      {\mbox{$P_T$}}
\newcommand {\abseta}  {\mbox{$\mid \eta \mid$}}

\newcommand {\sqrts}   {\mbox{$\sqrt{s}$}}
\newcommand {\BR}      {\mbox{$B$}}
\newcommand {\SrtB}    {\mbox{$S/\sqrt{B}$}}
\newcommand {\sigmaBR} {\mbox{$\sigma \times \BR$}}
\newcommand {\NPscale} {\mbox{$\Lambda$}}
\newcommand {\tanbeta} {\mbox{$\tan \beta$}}

\newcommand{\etal}       {\mbox{\it et al.}}

\begin{document}
%-----------------------------------------------------------------------
% Title Page
%-----------------------------------------------------------------------
\title{ HIGGS SEARCHES AT THE TEVATRON\protect\thanks{~~Talk given at 
	Les Rencontres de Physique de la Vall\'{e}e d'Aoste, 
	La Thuile, Italy; February 27 - March 4, 2000} }

\author{Harold G. Evans \\
	(for the CDF and \Dzero\ Collaborations) \\
      {\em Columbia University, Dept. of Physics } \\
	{\em 538 W. 120th St., MC5215 } \\
	{\em New York, NY 10027, U.S.A. }
       }

\maketitle
\baselineskip=14.5pt

\begin{abstract}
Higgs hunting is a world-wide sport and the Tevatron is set to become
the next field of play when Run II starts in March 2001.
To set the stage, we summarize results of searches for standard and
non-standard Higgs bosons by CDF 
and \Dzero\ in Run I at the Tevatron.
Progress has been made in quantifying the requirements on the
Tevatron Collider and on the upgraded experiments in Run II
for extending the excellent work done at LEP.
Armed with parameterizations of expected detector performance, the
Tevatron Higgs Working group has made predictions of the sensitivity
of CDF and \Dzero\ to Higgs bosons in the Standard Model and in its
Minimal Supersymmetric extension as a function of integrated luminosity.
These predictions are presented to underscore the excitement being
generated by Run II, and to highlight the need for the highest possible
luminosity.
\end{abstract}

\baselineskip=17pt
\newpage

%-----------------------------------------------------------------------
% Introduction
%-----------------------------------------------------------------------
\section{Introduction}
Despite increasingly precise scrutiny,
the standard model (SM) of particle physics remains largely unshaken.
Only the recent experimental evidence for massive neutrinos\cite{superk}
challenges any of its many predictions.
With the single exception of the Higgs boson,
all the particles expected in the SM, and no others, have now been
observed. 
This single missing particle represents a major gap in our knowledge of the
microscopic world. 
Tied up in its nature (or natures, in the case that there is more than
one Higgs scalar) is the method by which the original
\SUtwol$\times$\Uone\ symmetry of the theory is spontaneously broken
to the distinct electromagnetic and weak forces we observe.

Within the SM, the mass of the single physical Higgs boson left after
electroweak symmetry breaking is related to the
vacuum expectation value of the neutral Higgs field ($v$ = 246 GeV) by
$\MHsm ^2 = \lambda v^2$.
The Higgs self-coupling parameter, $\lambda$, is not specified by the
theory, making the Higgs mass an unknown quantity.
However, if the Higgs mechanism is to fulfill its role in the SM, the
Higgs mass cannot exceed 1 TeV,
otherwise unitarity would be violated in the scattering of
longitudinally polarized gauge bosons.

Since it does not incorporate gravity,
the SM cannot be a fundamental
theory of all interactions despite its many other successes.
Even if it is a correct effective theory up to the Planck scale of
$\sim$\unitexp{10}{19} GeV, where quantum gravitational effects become
significant, the SM is still unsatisfying.
This is because radiative corrections to the square of the Higgs mass are
quadratically divergent in an energy cutoff parameter, 
which should be the
Planck scale if the SM is to be valid up to that range. In order to get a
physical Higgs mass on the 
order of the electroweak scale, these corrections must be cancelled by
tuning the bare Higgs mass at the Planck scale to one part in
\unitexp{10}{16} -- a rather unnatural condition.

Since there is an implied need for new physics beyond the SM, a
natural question to ask is at what energy scale (\NPscale ) should such
effects become apparent.
The physics would then look SM-like below \NPscale , but  new
particles and interactions would become apparent beyond that scale.
This question is intimately related to the 
mass of the Higgs boson that is used to break electroweak symmetry.
At large values of \Mhlo , the renormalization group equation for the
Higgs self-coupling causes $\lambda$ to blow up for \NPscale\ below
the Planck mass. The point at which this happens then sets the scale
for new physics.
On the other hand, if \Mhlo\ is too small, top quark contributions to
$\lambda$ can drive it negative. To avoid this problem an energy cutoff must
be introduced that can be associated with \NPscale .
Using these constraints, a measurement of the Higgs mass at the
relatively low energies of today's accelerators could clarify
the scale for the breakdown of the SM.

Many theoretical frameworks have been proposed
to address some of the weaknesses of the SM
without abandoning its low energy successes.
These will not be discussed here.
We will limit our discussion to results of the minimal
supersymmetric standard model (MSSM)\cite{mssm},
a guide indicative of extensions 
that probe the nature of physics and the Higgs beyond the SM.

The MSSM is the simplest version of supersymmetric models that solve
the ``naturalness'' problem discussed above by relating fermionic and
bosonic degrees of freedom\cite{susy}.
This involves adding supersymmetric partners to all SM particles that
differ from their SM counterparts by a half unit of spin.
An additional requirement is the expansion of the Higgs sector to
include more than the single complex doublet used in the SM.
The MSSM proposes only a single extra complex
doublet. After electroweak symmetry breaking, five physical Higgs
particles remain:
two CP-even neutral scalars, \hlomssm\ and \Hhimssm\
(with \Mhlo $<$\MHhi\ by convention),
one CP-odd neutral scalar \Amssm\
and two charged scalars \Hpmmssm .

The parameters of the MSSM that most directly affect
the Higgs bosons are
\tanbeta\ (the ratio of the vacuum expectation values of the two Higgs
doublets) and
\MA\ (the mass of the CP-odd Higgs, 
conventionally chosen to be the free Higgs mass in the theory).
Unlike the case of the SM,
in the MSSM, the mass of the lightest of the Higgs bosons, \hlomssm ,
is constrained by supersymmetry. At tree level, this mass must be less
than that of the \Zz . Radiative corrections modify
this relationship, but an upper bound still exists with
(\Mhlo )$_{\mathrm{max}} \sim$ 130 GeV. 

The mass of an SM-like Higgs is also constrained by experimental
measurements. Direct searches at LEP 2 currently give the most
stringent lower bound on the mass. This bound changes as LEP
accumulates more data. A snapshot of the LEP-wide 95\% C.L. limit
given at the September 1999 LEPC\cite{mcnamara} constrains
\MHsm\ $>$ 102.6 GeV.
The best experimental upper bound on the Higgs mass comes from global
fits to electroweak measurements done by the LEP Electroweak Working
Group. Their results\cite{lepewwg99} indicate that
\MHsm\ $<$ 215 GeV at 95\% C.L.

%-----------------------------------------------------------------------
% The Higgs at the Tevatron
%-----------------------------------------------------------------------
\section{The Higgs at the Tevatron}
If a Higgs boson is not discovered at LEP 2 in its last year of
running, the next place to look will be the Tevatron at Fermilab.
The Tevatron is a \pbp\ collider that operated until
1996 at a center of mass energy of \sqrts\ = 1.8 TeV.
Two experiments, CDF\cite{cdfwww} and \Dzero\cite{d0www}, each collected
approximately 100 \unitexp{pb}{-1} of data during the period
1992--1996. 
As we will 
see, this data set, referred to as ``Run I'' data, is not sufficiently
large to have competitive sensitivity to an SM Higgs although
interesting studies have been made in certain non-standard models.

Not content with the success of Run I, the Tevatron accelerator
is being upgraded to increase its center-of-mass energy to \sqrts\ =
2.0 TeV and to ultimately achieve an instantaneous luminosity of
\scinot{2}{32} \unitexp{cm}{-2}\unitexp{s}{-1}\cite{tevupgr}.
CDF\cite{cdfupgr} and \Dzero\cite{d0upgr} are also being upgraded
to take advantage of the higher luminosity.
Data taking for this new ``Run II'' will start in March 2001.
Projections for integrated luminosity in Run II begin at 2
\unitexp{fb}{-1}, but could reach another factor of ten.
Obviously, with this large expected data set,
CDF and \Dzero\ have a bright future.

\begin{figure}
\begin{minipage}[t]{0.45\textwidth}
\begin{center}
  \mygraphics{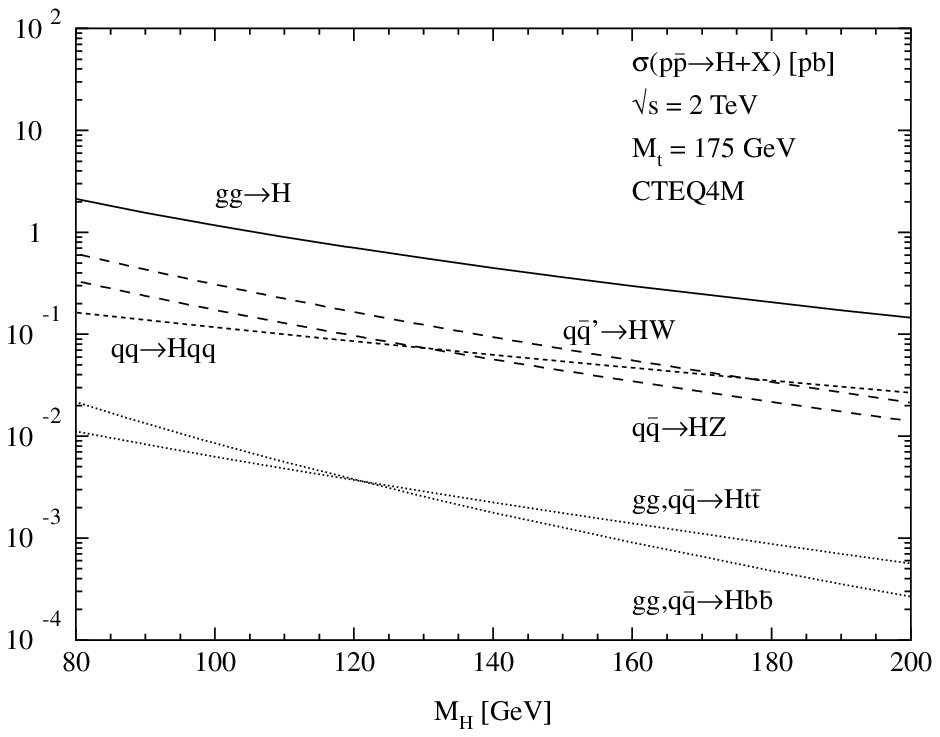}
  \caption{\it Production cross-sections in \ppb\ collisions at
	\sqrts\ = 2.0 TeV for various SM Higgs modes as a function of
	Higgs mass\cite{higgsprod}.
	\label{fig:hsmxs} }
\end{center}
\end{minipage}
\hfill
\begin{minipage}[t]{0.45\textwidth}
\begin{center}
  \mygraphics{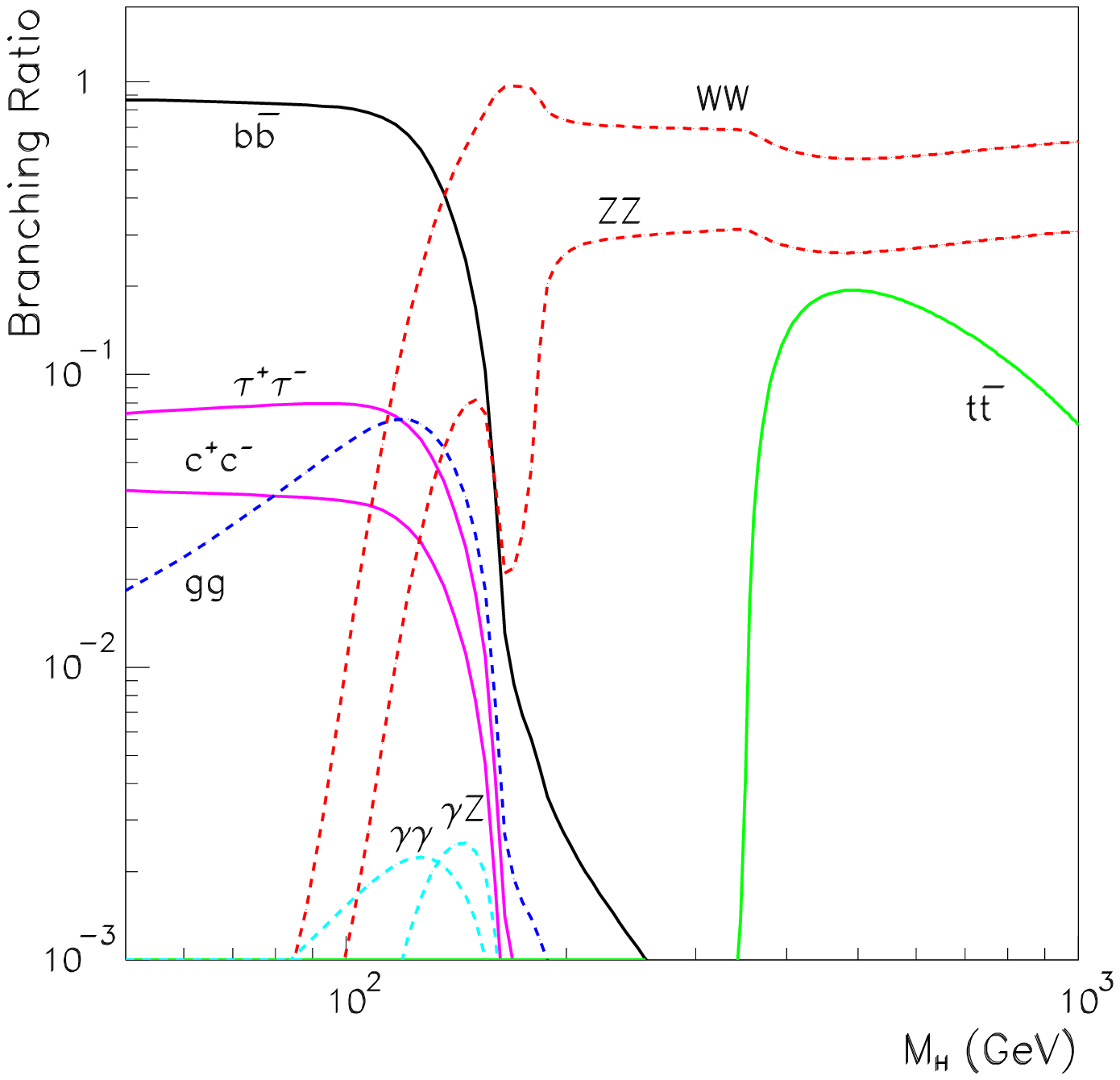}
  \caption{\it Branching ratios for the SM Higgs 
	as a function of Higgs mass\cite{fnalhwg}.
	\label{fig:hsmbr} }
\end{center}
\end{minipage}
\end{figure}

Before we dive into the details of Run I Higgs searches and
projections for Run II sensitivities, it is worthwhile to review how
Higgs bosons are produced and decay, and how they would be detected at
the Tevatron. 
The cross section for SM Higgs bosons in \pbp\ collisions,
as calculated by Spira\cite{higgsprod},
is given in Fig. \ref{fig:hsmxs} for various production modes.
The main SM Higgs decay modes, calculated using the program
HDECAY\cite{hdecay}, are given in Fig.
\ref{fig:hsmbr}\cite{fnalhwg} as a function of Higgs mass.
Corresponding plots for the neutral MSSM Higgs bosons, \hlomssm ,
\Hhimssm\ and \Amssm , also based on the calculations by Spira,
are available on the Fermilab Run II Higgs Working Group web page\cite{fnalhwg}.
These have largely similar characteristics to the plots for SM Higgs,
however, at large \tanbeta\ couplings to \qrk{b}-quarks and 
$\tau$-leptons are
enhanced, leading to \hlomssm (\Hhimssm )\bbb\ production being favored
over the \hlomssm (\Hhimssm)\Wboson /\Zboson\ modes.

The clear advantage of the Tevatron over \epem\ machines
is its center of mass energy --
a factor of 10 higher than at LEP.
Of course, not all of this energy is available to the partons
participating in the hard scattering producing the Higgs.
Nevertheless, higher mass Higgs bosons can be produced at the
upgraded Tevatron than at any previous machine.
The problem is finding them.

Table \ref{table:states} spells out the difficulty in identifying a
Higgs at the Tevatron -- the immense background.
Clearly, the most favorable production mode for the SM Higgs, \ggH\
(through intermediate top quarks),
cannot be used in a search for a low mass Higgs
(\MHsm $<$130 GeV),
that decays mainly to \bbb .
Background from QCD dijet production is a factor of $\sim$\unitexp{10}{6}
larger than the signal.
This has prompted Tevatron Higgs hunters to concentrate on the next
highest
cross section production modes of a light Higgs in association with a
\Wboson\ or \Zboson .
Even here, searches must contend with difficult background from
boson-pair, \Wboson /\Zboson \bbb\ and top quark production.

\begin{table}
\begin{center}
\caption{ \it Representative cross sections for SM Higgs production
	(\Mhlo\ = 100 GeV) and some major backgrounds\cite{fnalhwg}.
	\label{table:states} }
\vspace{0.1in}
\begin{tabular}{|c|c|}
\hline
  Mode & Cross-Section [pb] \\
\hline
  \ggH         & 1.0 \\
  \WH          & 0.30 \\
  \ZH          & 0.17 \\
\hline
  \WZ\ + \ZZ & 4.4 \\
  \Wboson \bbb\ + \Zboson \bbb       & 14 \\
\hline
  \ttb                                                    & 7.5 \\
  \qrk{t}\qrk{b} + \qrk{t}\qrk{q} + \qrk{t}\qrk{b}\qrk{q} & 3.4 \\
\hline
  QCD di-jet          & \order{\unitexp{10}{6}} \\
  QCD four-jet        & \order{\unitexp{10}{4}} \\
\hline
\end{tabular}
\end{center}
\end{table}

If the Higgs has higher mass, identification is slightly easier although the
production cross section is smaller because decays to \WW\
and \ZZ\ begin to dominate above \MHsm $\sim$130 GeV.
In fact, these unique final states, make it possible to
take advantage of the \ggH\ production mode in this mass region.

Regardless of the mass of the Higgs, searches at the Tevatron will
take advantage of all distinguishing features of Higgs decay
to overcome backgrounds with cross-sections many times
larger than those of the Higgs.
The most promising final states for a low mass SM Higgs 
(that decays mainly to \bbb ) are:
\lnbb , \llbb, \nnbb\ and \jjbb.
For a high mass SM Higgs (decaying mainly to \WW /\ZZ )
the best final states contain the following distinguishing particles:
\llnn , \lpmlpmjj\ and \lpmlpmlmp .
Discussion of the motivation for choosing these final states can be
found in Run II Higgs Working Group Report\cite{fnalhwg}.

A few themes emerge from consideration of Table \ref{table:states}
and the final states listed above.
To be maximally sensitive to the Higgs, an ideal Tevatron detector must
have all of the following properties:
\begin{enumerate}
  \item High lepton (e and $\mu$) identification efficiency,
  \item Excellent missing energy resolution (for neutrino
  identification),
  \item High \qrk{b}-quark identification efficiency,
  \item Good invariant mass resolution for \bbb\ pairs (to reject
  \bbb\ background outside of the \Hbb\ peak).
\end{enumerate}

The same signatures mentioned for the SM Higgs will also be
important for finding neutral Higgs scalars in the MSSM.
Since couplings to \qrk{b}-quarks and $\tau$-leptons tend to grow with
increasing \tanbeta , $\tau$ identification takes on greater
importance and good sensitivity to \qrk{b} jets
becomes even more crucial.

An MSSM charged Higgs with mass less than $m_t - m_b$ is
expected to be produced at the Tevatron through the decay of top
quarks --
a very different mechanism than for neutral Higgs particles.
In regions of large and small \tanbeta\ 
(away from \tanbeta $\sim$6)
the branching ratio for
\qrk{t} \ra \Hpmmssm \qrk{b} is predicted to be quite large. The decay
of the \Hpmmssm\ is mainly to \Wboson \bbb\ and \qrk{c}\qrk{s} for
low \tanbeta\ and to $\tau$\nutau\ for high \tanbeta .
These final states are sufficiently different from the SM top-quark decays
that standard top analyses would have low efficiency for them.
This means that,
in addition to looking explicitly for the \Hpmmssm\ decay products
(especially $\tau\nu$)
a ``disappearance'' search is also possible for \Hpmmssm .
For a substantial branching ratio of top to charged Higgs,
the measured \ppb \ra \ttb $X$ cross-section would be smaller than the
SM expectation.
A discrepancy between measurement and prediction can provide evidence for
\Hpmmssm .

%-----------------------------------------------------------------------
% Higgs Searches at Run I
%-----------------------------------------------------------------------
\section{Higgs Searches at Run I}
Searches for Higgs have been performed by CDF and \Dzero\ in all the main SM
final states
as well as in models beyond the SM.
Results are summarized in Table \ref{table:runi}.
The reach of CDF and \Dzero\ in production cross-section
multiplied by \Hbb\ branching ratio in standard Higgs modes
is far weaker than the predictions of the SM. For
certain models beyond the SM, however, sizable regions of parameter
space can be excluded.
The charged Higgs of the MSSM is an especially interesting search
since the Tevatron is the only facility that can take advantage of
the \tHb\ mode.

\begin{table}
\begin{center}
\caption{ \it A summary of results of Run I Higgs searches by CDF
	and \Dzero .
	Production cross-section multiplied by the
	\Hbb\ branching ratio (\sigmaBR ) limits are for a 100 GeV Higgs,
	where \BR $\sim$0.81.
	Values marked with an asterisk are preliminary.
	\label{table:runi} }
\vspace{0.1in}
\begin{tabular}{|l|l|l|l|l|}
\hline
  & Contributing  & Predicted       & \multicolumn{2}{|c|}{95\% C.L. Limits} \\
    \cline{4-5}
  Channel & Modes & \sigmaBR\ [pb]  & CDF               & \Dzero \\
\hline
  \multicolumn{5}{|r|}{\sigmaBR\ Limits [pb]} \\
\hline
  \lnbb   & \WH        & 0.24 & $<$27\cite{cdfcomb} & * $<$28\cite{d0lnbb} \\
  \jjbb   & \WH +\ZH   & 0.38 & $<$23\cite{cdfcomb} & --- \\
  comb.   & \WH +\ZH   & 0.38 & $<$17\cite{cdfcomb} & --- \\
  & & & & \\
  \nnbb   & \ZH        & 0.14    & * $<$8  & * $<$\order{70}\cite{d0nnbb} \\
  \llbb   & \ZH        &         & * $<$38             & --- \\
  comb.   & \ZH        &         & * $<$7.5            & --- \\
  & & & & \\
  \bbb \bbb & \phibb & 0.01$\cdot \tan^2 \beta$
                                 & * $<$26 pb             & --- \\
\hline
  \multicolumn{5}{|r|}{\Mhlo\ Limits [GeV]} \\
\hline
  $\gamma \gamma X$ & \hlomssm \ra $\gamma \gamma$
                       & 0.07 & * $>$82\cite{cdfhgg} & $>$78.5\cite{d0hgg} \\
\hline
  \multicolumn{5}{|r|}{\BR (\tHb ) Limits} \\
\hline
  \ttb -disappearance           & \tHb
    & & * $<$32\%\cite{bevensee} & $<$45\%\cite{d0hpm} \\
  $(\ell \ell) / (\ell +j)$   & \tHb \ra \qrk{c}\qrk{s}\qrk{b}
    & & * $<$72\%\cite{bevensee} & --- \\
  $\tau \ell j$                 & \tHb \ra $\tau \nu$\qrk{b}
    & & $<$60\%\cite{cdfhpmlt} & --- \\
\hline
  \multicolumn{5}{|r|}{\MHpm\ Limit [GeV]} \\
\hline
  $\tau jjX$+2$\tau$               & \tHb \ra $\tau \nu$\qrk{b}
    & & $>$147\cite{cdfhpmht} & --- \\
\hline
\end{tabular}
\end{center}
\end{table}

%--------------------------------
% The Four b Final State at CDF
%--------------------------------
\subsection{The Four-\qrk{b} Final State at CDF}
A preliminary analysis from CDF is another good example of the
possibilities of Run I searches in beyond-the-SM scenarios.
Here, \qrk{b}-quark identification of the CDF detector is used to
select events with four \qrk{b} jets in the final state.
This topology is expected at large \tanbeta\ in several SUSY models where
the coupling of some of the neutral Higgs scalars to \qrk{b}-quarks is
enhanced\cite{thbbbb}.
The final state arises in \bbb\ production when one of the primary
\qrk{b}-quarks radiates a neutral Higgs which, then decays to \bbb .
The cross section for this process goes as $\tan^2 \beta$, and can
therefore become sizable at large \tanbeta .
The cross section is also affected by details of the mixing between
left- and right-handed stop squarks. Results are presented for two
extreme cases: that of minimal and maximal stop mixing.

This CDF analysis,
which is an update of that presented in the Run II Higgs Working Group
Report\cite{fnalhwg}, 
uses a multijet trigger requiring a
total cluster energy of $>$125 GeV and at least four trigger
clusters with energy $>$15 GeV for
an integrated luminosity of 91 \unitexp{pb}{-1}.
Offline, at least four jets are required, with \Et $\geq$15 GeV and 
\abseta $<$2.4\footnote{The pseudo-rapidity, $\eta$, is defined as
  $\eta = -\ln(\tan \frac{\theta}{2})$, where $\theta$ is the polar angle.}.
Three of these jets must be tagged
by the CDF secondary vertex algorithm\cite{secvtx}
as arising from \qrk{b}-quarks.
In order to further reject the large QCD multi-jet background,
additional criteria are imposed to take advantage of the distinct topology of
four \qrk{b} jets arising from \phibb\ 
(\Hgeneral\ = \hlomssm ,\Hhimssm ,\Amssm ) production.
These criteria involve
an $M_{\Hgeneral}$ dependent cut on the \Et\ of the three highest-\Et\
jets,
a requirement that the azimuthal angle between the two leading \qrk{b}
jets
be larger than 1.9 radians,
and an $M_{\Hgeneral}$ dependent cut on the invariant mass
of several different jet pairings.

\begin{figure}
\begin{center}
  \mygraphics{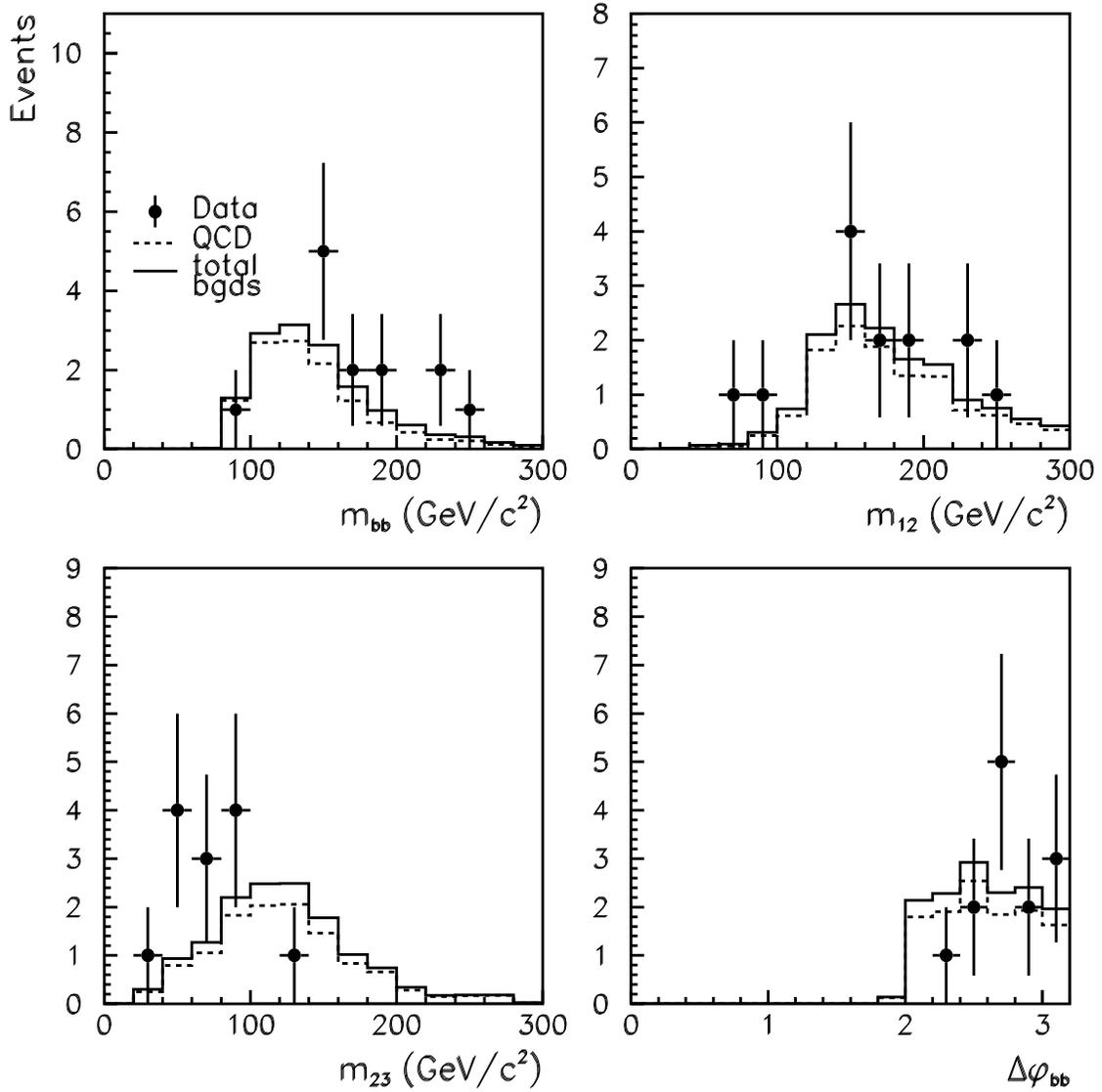}
  \caption{\it Distributions of data in the 3-\qrk{b} selected sample
	for the largest invariant mass \qrk{b}-tagged jet pairs
	($m_{bb}$), the mass 
	combination of the highest-\Et\ and next-highest-\Et\ jets
	($m_{12}$), 
	the mass of the second- and third-highest \Et\ jets ($m_{23}$)
	and the azimuthal difference between the two leading
	\qrk{b}-tagged jets ($\Delta \varphi_{bb}$).
	\label{fig:4bdata} }
\end{center}
\end{figure}

\begin{figure}
\begin{minipage}{0.45\textwidth}
\begin{center}
  \mygraphics{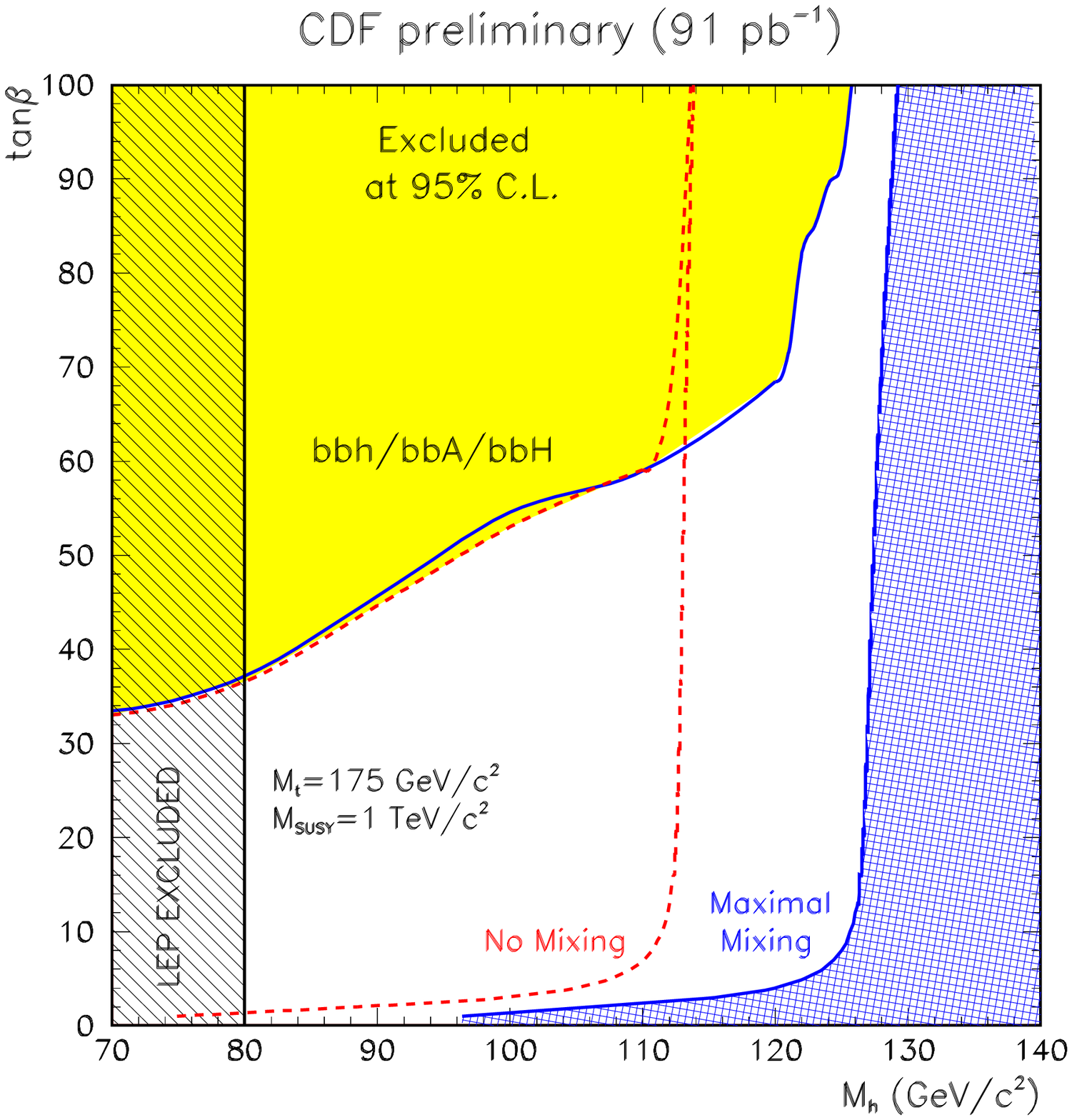}
\end{center}
\end{minipage}
\hfill
\begin{minipage}{0.45\textwidth}
\begin{center}
  \mygraphics{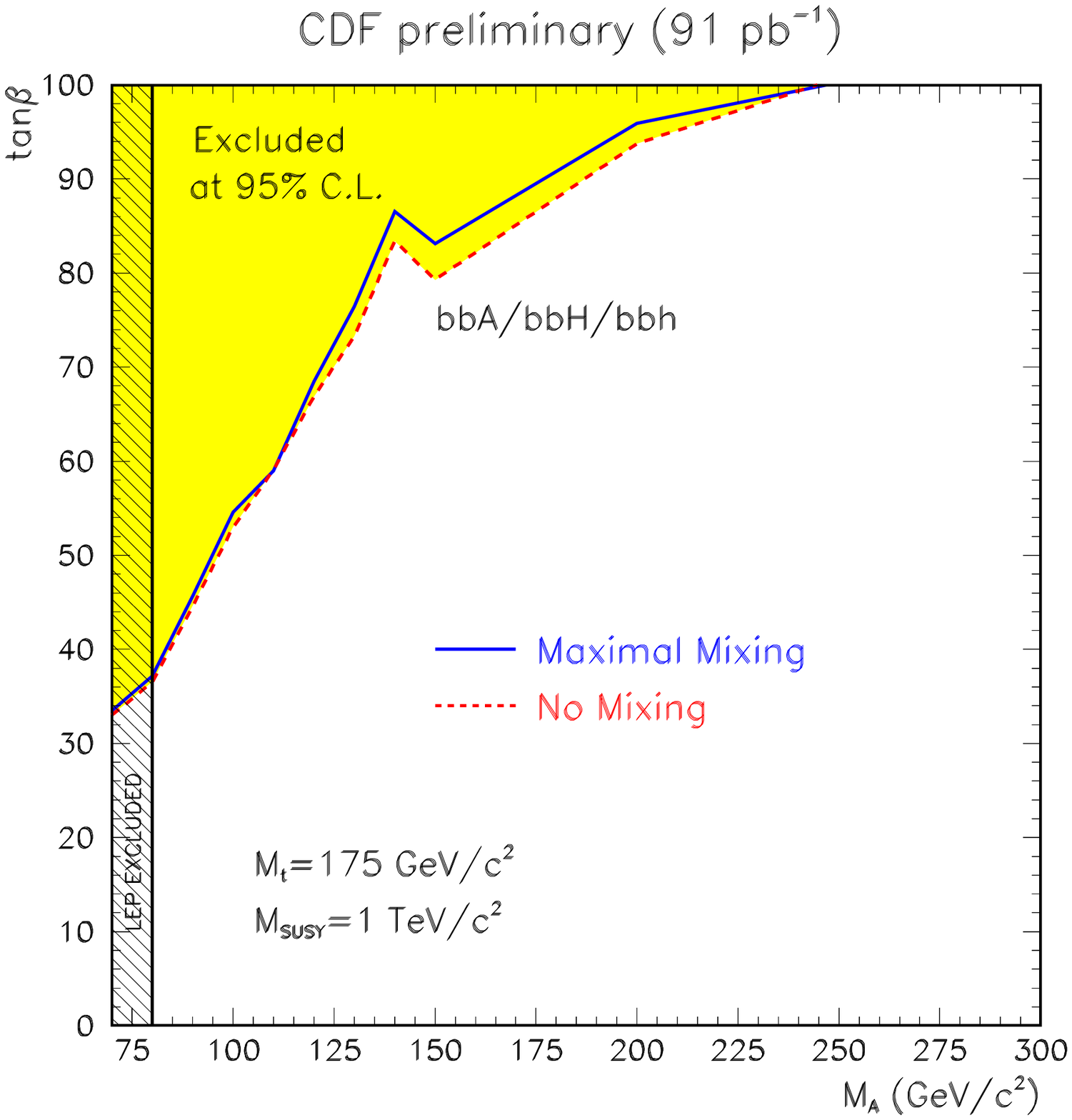}
\end{center}
\end{minipage}
\begin{center}
  \caption{\it The region excluded at 95\% C.L. by the CDF \phibb\
	analysis in \tanbeta\ vs. \Mhlo\ (left plot) 
	and \MA\ (right plot) parameter space.
	The double-hatched region corresponds to theoretically
	forbidden values of \Mhlo .
	The solid line corresponds to the case of no stop-squark
	mixing, and the dashed line to maximal mixing.
	\label{fig:4bres} }
\end{center}
\end{figure}

Distributions in several variables used in the analysis, comparing
data and SM prediction, are shown in Fig. \ref{fig:4bdata} for CDF data
with three \qrk{b}-tagged jets,
prior to the imposition of the mass requirements.
A breakdown of the efficiencies and backgrounds in the analysis, along
with the number of observed events in the final selection is given in
Table \ref{table:4bres}.
No evidence for a \phibb\ signal is seen, providing a 95\% C.L. limit
on \sigmaBR\ for this process of 25.7 pb.
This limit can be interpreted within the framework of the MSSM as an
exclusion region in \tanbeta\ vs. \Mhlo\ or \MA\ space.
The excluded regions are presented in Fig. \ref{fig:4bres}.
As can be seen, the excluded regions extend significantly those
previously probed by LEP.

\begin{table}
\begin{center}
\caption{ \it Results of the preliminary CDF \phibb\ search
	including efficiencies at various stages of the selection
	process and
	the number of background and observed events in the final selection.
	Backgrounds are calculated for the $M_{\Hgeneral}$=100 GeV
	selection and for the integrated luminosity of the data sample.
	\label{table:4bres} }
\vspace{0.1in}
\begin{tabular}{|rr|rr|}
\hline
  \multicolumn{2}{|l|}{Efficiencies} & \multicolumn{2}{|l|}{Backgrounds} \\
  Trigger & 1.9\% & & \\
  Jet \Et & 86\% & QCD  & 2.2$\pm$1.1 \\
  $\geq$3 \qrk{b}'s & 20\% & Fakes & 0.5$\pm$1.4 \\
  $\Delta \varphi$\qrk{b}\qrk{b} & 82\% & \Wboson \bbb /\ccb & 0.1$\pm$0.1 \\
  Mass & 93\% & \Zboson \bbb /\ccb & 0.37$\pm$0.02 \\
    Total Effic. & 0.25\% & Total Bgrd. & 3.8$\pm$1.1 \\
\hline
  Observed Events & 3 
  & \sigmaBR\ Limit & 25.7 pb \\
\hline
\end{tabular}
\end{center}
\end{table}

%-----------------------------------------------------------------------
% Higgs Searches at Run II
%-----------------------------------------------------------------------
\section{Higgs Searches at Run II}
Despite the good sensitivity of Run I analyses to several non-SM
Higgs bosons, the results for the SM Higgs are (as expected) not
significant. As we saw in Table \ref{table:runi}, \sigmaBR\ limits
for final states expected from the SM Higgs are, at best, a factor of 50
higher than predictions.
Given this picture, why are we optimistic about Higgs searches
in Run II?
There are three reasons: higher \sqrts , higher luminosity 
and better detectors.

The increase in center of mass energy of the Tevatron for Run
II is modest (1.8 to 2.0 TeV). 
This translates, however, into a substantial gain in cross section for
associated Higgs production of approximately 20\% in the SM.

As mentioned previously, the integrated luminosity collected in Run II
is expected to be at least 2 \unitexp{fb}{-1} by 2002 
and should reach $\sim$15 \unitexp{fb}{-1} before the start of the LHC.
This large data set
will be our main lever on the SM Higgs.
However, it should not be forgotten that if we see evidence
for a Higgs particle, larger 
control samples such as \Zboson \ra \bbb\ that will be
available in Run II will give us confidence that what we observe
actually corresponds to signal.

In order to take maximum advantage of the glorious new data sets, both
CDF\cite{cdfupgr} and \Dzero\cite{d0upgr} are being upgraded.
From experience with Run I analyses and some theoretical guidance, a
clear picture has emerged of the most important detector properties
required for Higgs searches.
Missing energy, leptons and \qrk{b}-quarks are the experimental
pillars of Higgs searches at hadron machines.
Efficient identification and accurate reconstruction of these objects
requires all features of the detectors to work at full capacity, and
consequently all aspects are being overhauled.
Some of the improvements that have the most impact on Higgs searches
are mentioned below.

Improving lepton identification is mainly a question of increasing the
coverage of the calorimeters for electrons 
(which also determines missing energy resolution)
and the muon chambers for muons.
The Run I calorimeters of both experiments were excellent.
Therefore, no changes are being made,
aside from those required to adapt to the new beam conditions.
CDF and \Dzero\ are, however, both increasing the effective coverage
of their muon systems.

Identification and reconstruction of \qrk{b}-quarks depends critically on
tracking (although soft lepton identification also plays a role).
To improve prospects,
the old CDF silicon detector
is being replaced 
with new 3D readout detectors that provide
stand-alone silicon tracking to \abseta $<$2.0.
This amounts to a 40\% increase in acceptance.
Using this detector, CDF expects to gain in the efficiency of double
\qrk{b}-tagging for \ttb\ events by a factor of 3.5\cite{layer00}. 
The tracking system of \Dzero\ will be even more radically revamped.
Central to this is the addition of a magnet providing a solenoidal field of
2.0 T in the tracking volume. The tracking system will consist
of cylinders of scintillating fibers and a silicon detector with 3D
readout extending to \abseta $<$1.7. 
This will allow \Dzero\ to join in
the \qrk{b}-quark game at the same level as CDF.
Both detectors are also adding dedicated trigger systems to identify
\qrk{b}-quarks online.

%---------------------------
% Prophecies for Run II
%---------------------------
\subsection{Prophecies for Run II}
We now embark into the realm of speculation
about what will happen in Run II.
This is not purely an exercise in fantasy, because
it is extremely important to understand how detector limitations will
affect Higgs searches while these parameters can still be
modified. It is also crucial to know what luminosity is required to
achieve sensitivity to the Higgs at different masses, as this will
strongly influence the running strategy.
As such, a Fermilab-wide working group, 
consisting of representatives from CDF, \Dzero\ and
the Theory Group,
was established to study Higgs issues at Run II.
The results presented in this section are based mainly on a
preliminary version of the Working Group report (version 3)
available at the time of the conference.
The most up to date version (version 6) can be found on the Working
Group's web page\cite{fnalhwg}.

Of course, to make predictions that have any chance of correctly
fortelling the future we need accurate simulations of key performance
parameters. Unfortunately, full simulations of the upgraded CDF and
\Dzero\ detectors are still evolving. However,
even with a relatively simple simulation, using parameterized detector
response, 
we can go a long way
towards answering questions that are relevant for the construction and
early running phases of Run II concerning
detector resolution and efficiencies required for Higgs sensitivity
and luminosity limitations on the mass reach.

The simulation used for the bulk of the results presented here,
referred to as SHW, parameterizes important detector resolutions and
efficiecies using an ``average'' of the foreseen Run II CDF and
\Dzero\ detectors\cite{fnalhwg}.
Most of the detector parameters in SHW are tunable, allowing studies
to be made of how a specific parameter impacts Higgs sensitivity.
As a baseline, most analyses use
a track reconstruction efficiency of 97\% for tracks with \abseta $<$2
and \Pt $>$300 MeV,
relative energy resolutions for the electromagnetic and hadronic calorimeters
of 20\%/$\sqrt{E}$ and 80\%/$\sqrt{E}$ respectively,
\qrk{b}-tagging efficiency of $\sim$60\% for \Et =100 GeV
and a \bbb\ relative mass resolution of 10--14\%.

A few warnings about SHW are in order.
First, since SHW is a parameterized simulation, details of
event-by-event detector response are missing.
This means that systematic effects and hardware-related background and
misidentifications are largely neglected. Their impact can be
estimated, however, from 
extrapolations of Run I results, and thus are not completely ignored.
Another difficult issue concerns the trigger.
Excellent trigger performance will be crucial to obtaining good
results.
The most questionable of these, the hadronic event triggers, are
considered in the
analyses outlined here, however, leptonic triggers are generally taken
to be 100\% efficient. This is a reasonable assumption if lepton
triggers function as foreseen.
In general, SHW predictions should be taken in their context --
a means to understand the detector and accelerator requirements so as to
achieve competitive sensitivity to Higgs.
While more elaborate simulations may yield slightly more accurate
predictions in some areas, only data will tell us the real story.

Before turning to channel-by-channel sensitivities as a function of
Higgs mass and luminosity it is worthwhile to describe the SHW results
concerning 
the key Higgs-search detector parameters:
missing energy resolution, lepton identification, \qrk{b}-quark
identification and \bbb\ mass resolution.
\begin{enumerate}
  \item Missing \Et\ resolution was excellent in Run I and no gains
        are foreseen in this area.
  \item Lepton identification efficiency in CDF and \Dzero\ is mainly
        governed by geometrical acceptance. Improvements are being
        made in the muon systems of both detectors.
  \item Tagging of \qrk{b}-quarks plays an important role in
        any Higgs search. However, signal significance
        (\SrtB ) grows at a faster rate if \bbb\ mass resolution
        is improved than if \qrk{b}-tagging efficiency is
        increased. This highlights the importance of a good
        understanding of the \qrk{b}-jet energy scale.
\end{enumerate}

\begin{figure}[t]
\begin{minipage}[t]{0.45\textwidth}
\begin{center}
  \mygraphics{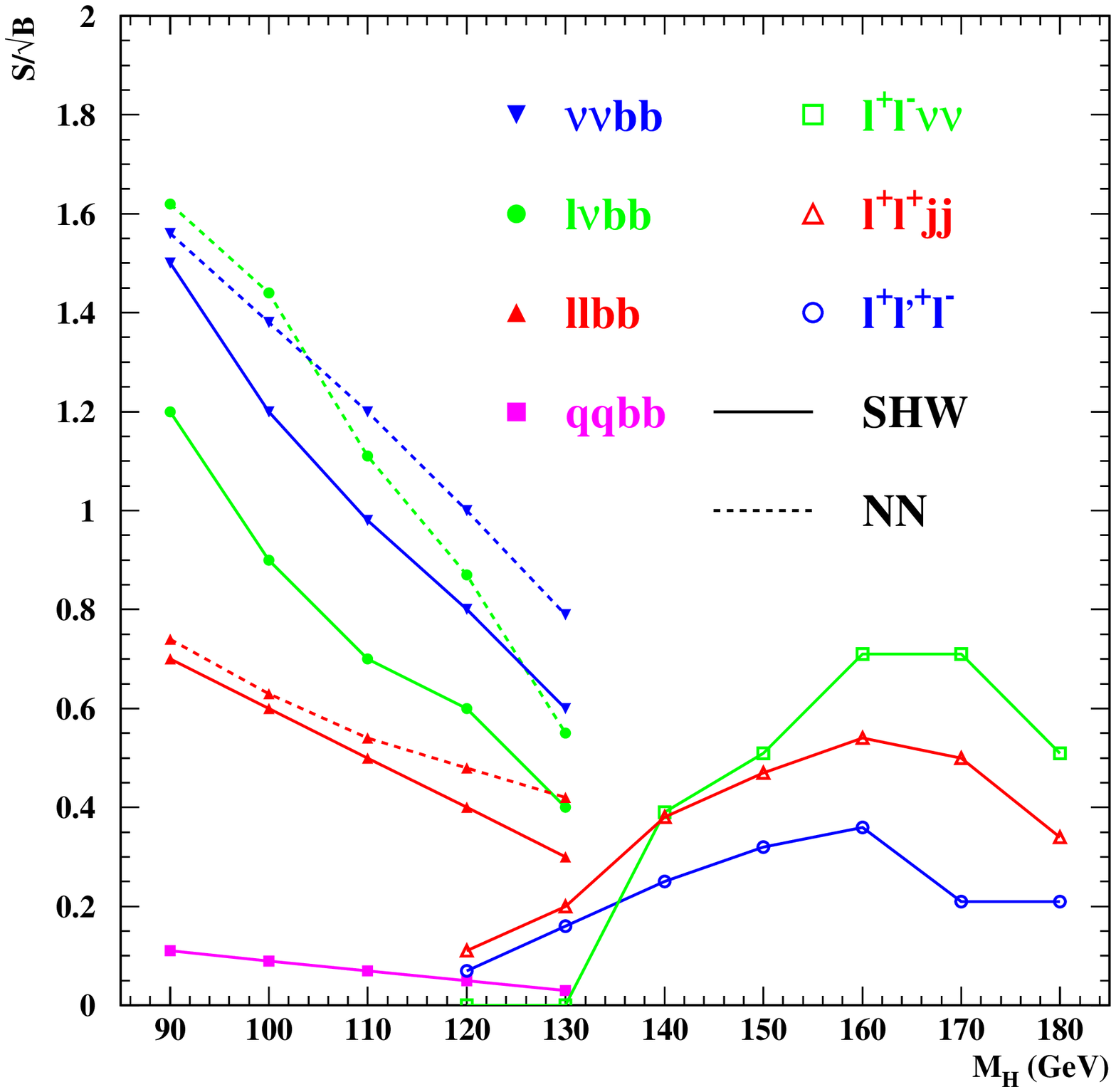}
  \caption{\it Sensitivities for a combination of CDF and \Dzero\
	expectations in the main SM Higgs final states. Results are
	given for 1 \unitexp{fb}{-1} delivered to each experiment.
	Points connected by solid lines correspond to cut-based
	analyses, while the dashed lines indicate results using a
	neural nets.
	\label{fig:sens} }
\end{center}
\end{minipage}
\hfill
\begin{minipage}[t]{0.45\textwidth}
\begin{center}
  \mygraphics{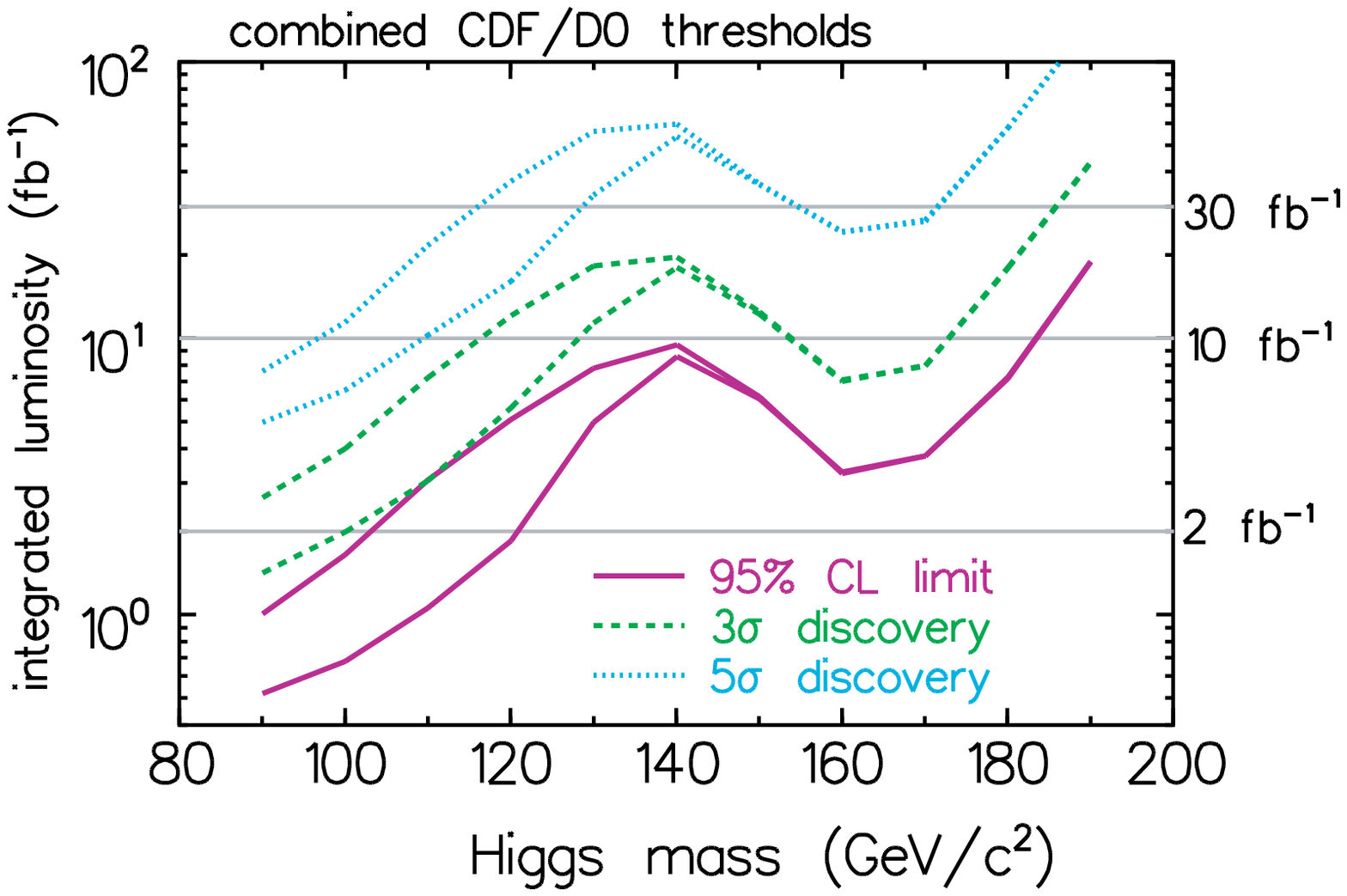}
  \caption{\it SM Higgs sensitivity predicted for a combination of CDF
	and \Dzero\ expectations as a function of Higgs mass and
	integrated luminosity delivered to each experiment.
	Limits (at 95\% C.L.), 3$\sigma$ evidence and 5$\sigma$
	discovery curves are plotted for cut-based (upper lines) and
	neural net-based (lower lines) analyses.
	\label{fig:mvsl} }
\end{center}
\end{minipage}
\end{figure}

Sensitivity to an SM Higgs boson from a combination of CDF and \Dzero\
expectations, 
as measured by signal over the square-root of background (\SrtB ),
for various decay channels, as a function of Higgs mass, is presented in
Fig. \ref{fig:sens} for an integrated luminosity of 1
\unitexp{fb}{-1} per experiment.
Several points are apparent.
First, the mass reach of the Tevatron experiments is significantly
improved by considering final states produced when the Higgs 
(at high mass)
decays to real or virtual boson pairs.
Second, good improvements in sensitivity over purely cut-based
analyses can be expected when using multi-variate techniques such as
neural net analyses\cite{nnanal}.
Finally, it is clear that 1 \unitexp{fb}{-1} per experiment will not
get us to the Higgs.
This is quantified in Fig. \ref{fig:mvsl} where the
combined CDF and \Dzero\ 95\% C.L. limits, 3$\sigma$ evidence and
5$\sigma$ discovery thresholds for a given integrated luminosity
delivered to each experiment are plotted as a function of Higgs mass.
With the minimal Run II integrated luminosity of 2 \unitexp{fb}{-1},
the Tevatron will barely, if at all, extend the expected LEP2 Higgs
mass limit of $\sim$115 GeV\cite{eilam}.
With more than 10 \unitexp{fb}{-1} per experiment, an SM Higgs could
be excluded up to around 180 GeV.
First hints of a real Higgs signal (at 3$\sigma$) would only appear
beyond what has 
already been excluded for an integrated luminosity of at least 20
\unitexp{fb}{-1} per experiment.

An important consideration in making the projections in Fig.
\ref{fig:mvsl} is our confidence in the predictions for background.
Estimates of background levels based purely on Monte Carlo are
notoriously unreliable;
especially those originating from tails of distributions.
An example is the QCD \bbb\ background in the \nnbb\ channel.
To take account of this unreliability, a relative uncertainty on the
background ($B$) in each channel of a minimum of 10\% or
1/$\sqrt{LB}$ (where $L$ 
is the integrated luminosity) is used in combining the individual
channels to produce Fig. \ref{fig:mvsl}.
Of course, the increased luminosity of Run II 
should provide better understanding of the background
based on control data
samples, which can be used tighten selection criteria,
thereby reducing background systematics.

Projected sensitivities for both neutral and charged Higges in the
framework of the MSSM\cite{fnalhwg}
indicate that
a relatively high integrated luminosity
(10--15 \unitexp{fb}{-1} for exclusion or
20--30 \unitexp{fb}{-1} for discovery)
will also be needed to reach decisive conclusions.
However, if this is delivered, SUSY could be discovered or constrained
over significant regions of the MSSM parameter space.

%-----------------------------------------------------------------------
% Conclusions
%-----------------------------------------------------------------------
\section{Conclusions}
As we have seen, the Tevatron has been a very active field for Higgs
searches. Several interesting limits have come out of Run I analyses
relevant to predictions beyond the SM,
but results for the minimal Higgs have not dented the SM.
Nevertheless, the valuable experience gained in Run I, is already
being applied to 
the upcoming Run II, slated to start in March of 2001.
In order to have the best possible detectors for Higgs searches and to
help set optimal parameters for the next run, studies have been
initiated by the
Fermilab Run II Higgs Working Group to determine the effect of detector
choices and luminosity on the Higgs reach at the Tevatron.
The main improvement in Run II that makes us optimistic about Higgs
prospects is certainly the increased luminosity, a factor of 20 or
perhaps as much as 300 over that delivered in Run I.
Detector improvements will also play a big role. The main gains here
come in \qrk{b} identification and \bbb\ mass resolution -- both of
which are essential for discovering the Higgs.
Not to be overlooked is the fact that \Dzero\ will fully enter the
arena of \qrk{b}-tagging with their upgraded tracking system in Run
II. This will have a major impact on the overall Tevatron sensitivity.
The big lesson that Run II Higgs prophecies teach us though is that
luminosity will be crucial. With only the initial 2 \unitexp{fb}{-1} the
Tevatron will not extend the eventual LEP2 Higgs sensitivity, of
$\sim$115 GeV\cite{eilam}.
With 10 \unitexp{fb}{-1} we could exclude at 95\% C.L. an SM Higgs up
to $\sim$180 GeV
and with 20 \unitexp{fb}{-1} we could see evidence at the 3$\sigma$
level for Higgs masses up to 180 GeV.
Similarly, strong sensitivity to a wide region of MSSM parameter space
can be made with more than 10 \unitexp{fb}{-1}.
These sensitivities are especially interesting given that the lightest
Higgs is predicted to lie below $\sim$130 GeV in the MSSM.
Needless to say, anticipation at the Tevatron is high!

%-----------------------------------------------------------------------
% Acknowledgements
%-----------------------------------------------------------------------
\section{Acknowledgements}
This talk would have been entirely free of content without the help of
a large number of people. I would especially like to acknowledge the
wise advice of
Max Chertok, Regina Demina, Mark Kruse, Andr\'{e} Turcot, Juan Valls
and Weiming Yao. Finally, huge thanks go to the conference
organizers for providing us with such a stimulating meeting and such a
good snow fall.

%-----------------------------------------------------------------------
% References
%-----------------------------------------------------------------------


\section{References}
\begin{thebibliography}{99}
\bibitem{superk}
	Y.Fukuda, \etal\ (SuperKamiokande),
	\Journal{\PRL}{81}{1562}{1998}.
\bibitem{mssm}
	see for example, S.Dawson
	``SUSY and Such'',
	Lectures given at the
	NATO Advanced Study Institute on Techniques and Concepts 
	of High-energy Physics, St. Croix, U.S. (1996),
	\hepref{ph}{9612229},
	and references therein.
\bibitem{susy}
	see for example, H.P.Nilles,
	\Journal{Phys. Rep.}{110}{1}{1984}.
\bibitem{mcnamara}
	P.McNamara, talk given at the LEPC meeting, September 7, 1999.
\bibitem{lepewwg99}
	D.Abbaneo, \etal\ (LEP Electroweak W.G.)
	\CERNEP{2000}{016}.
\bibitem{cdfwww}
	A detailed description of the CDF detector and physics results
	can be found in the web pages,
	\WWWAddr{http://www-cdf.fnal.gov/}
\bibitem{d0www}
	A detailed description of the \Dzero\ detector and physics results
	can be found in the web pages, 
	\WWWAddr{http://www-d0.fnal.gov/}
\bibitem{tevupgr}
	\WWWAddr{http://www-bd.fnal.gov/lug/},
	see especially ``The Run II Handbook'',
	\WWWAddr{http://www-bd.fnal.gov/lug/runII\_handbook/RunII\_index.html}
\bibitem{cdfupgr}
	\WWWAddr{http://www-cdf.fnal.gov/}
	see especially ``The CDF II Detector Technical Design Report'',
	Fermilab-Pub-96/390-E,
	\WWWAddr{http://www-cdf.fnal.gov/upgrades/tdr/tdr.html}
\bibitem{d0upgr}
	\WWWAddr{http://www-d0.fnal.gov/}
	see especially 
	``The \Dzero\ Upgrade: The Detector and its Physics''
	Fermilab-Pub-96/357-E,
	\WWWAddr{http://higgs.physics.lsa.umich.edu/dzero/d0doc96/d0doc.html},
\bibitem{higgsprod}
	M.Spira,
	\hepref{ph}{9810289}.
\bibitem{hdecay}
	A.Djouadi, J.Kalinowski and M.Spira,
	\Journal{Comput. Phys. Commun.}{108}{56}{1998}.
\bibitem{fnalhwg}
	``Report of the Higgs Working Group'' from the Physics at Run
	II Supersymmetry/Higgs Workshop,
	\WWWAddr{http://fnth37.fnal.gov/susy.html}. \\
	see also \WWWAddr{http://fnth37.fnal.gov/higgs.html} \\
	Note that results presented here are from version 3 (Sep., 29,
	1999) of the report. The current version is 6.
\bibitem{cdfcomb}
	F.Abe \etal\ (CDF),
	\Journal{\PRL}{81}{5748}{1998}.
\bibitem{d0lnbb}
	S.Abachi, \etal\ (\Dzero ),
	``Results from a Search for a Neutral Scalar Produced in
	Association with a \Wboson\ Boson in \ppb\ Collisions at
	\sqrts\ = 1.8 TeV'',
	contributed paper to the
	{\it 28th International Conference on High-Energy Physics
	 (ICHEP-96)}, Warsaw, Poland (1996),
	\FermiConf{96}{258}.
\bibitem{d0nnbb}
	B.Abbott, \etal\ (\Dzero ),
	``Search for \Zboson $X$\ra \nnbb\ Events in the \Dzero\ Detector'',
	contributed paper to the
	{\it XVIIIth International Symposium on Lepton Photon Interactions},
	 Hamburg, Germany (1997).
\bibitem{cdfhgg}
	P.J.Wilson (for CDF),
	``Search for High Mass Photon Pairs in \ppb\ Collisions at
	\sqrts\ = 1.8 TeV'',
	contributed paper to the
	{\it 29th International Conference on High-Energy Physics
	 (ICHEP-98}, Vancouver, British Columbia (1998),
	\FermiConf{98}{213}..
\bibitem{d0hgg}
	B.Abbott, \etal\ (\Dzero ),
	\Journal{\PRL}{85}{2244}{1999}.
\bibitem{bevensee}
	B.Bevensee (for CDF),
	proceedings of the
	{\it 33rd Rencontres de Moriond, QCD and Hadronic
	Interactions},
	Les Arcs, France (1998),
	\FermiConf{98}{155}.
\bibitem{d0hpm}
	B.Abbott, \etal\ (\Dzero ),
	\Journal{\PRL}{82}{4975}{1999}.
\bibitem{cdfhpmlt}
	T.Affolder, \etal\ (CDF),
	submitted to \PRD ,
	\hepref{ex}{9912013}.
\bibitem{cdfhpmht}
	F.Abe, \etal\ (CDF),
	\Journal{\PRL}{79}{357}{1997}.
\bibitem{thbbbb}
	see for example,
	D.Dicus, T.Stelzer, Z.Sullivan and S.Willenbrock,
	\hepref{ph}{9811492}.
\bibitem{secvtx}
	F.Abe, \etal\ (CDF),
	\Journal{\PRL}{74}{2626}{1995}.
\bibitem{layer00}
	``Proposal for Enhancement of the CDF II Detector'',
	presented to the Fermilab Director and PAC,
	October 23, 1998,
	Fermilab-Proposal-909.
\bibitem{nnanal}
	P.Bhat, R.Gilmartin and H.Prosper,
	\hepref{ph}{0001152}.
\bibitem{eilam}
	E.Gross,
	these proceedings. \\
	see also E.Gross, A.L.Read, D.Lellouch
	\CERNEP{98}{094}.
\end{thebibliography}
\end{document}